# Enhancement of high dielectric constant in $CaCu_3Ti_4O_{12}/RuO_2$ composites in the vicinity of the percolation threshold.

Rupam Mukherjee, Gavin Lawes, and Boris Nadgorny

Department of Physics and Astronomy, Wayne State University, Detroit, Michigan, 48201

## Abstract

We observe the large enhancement in the dielectric response near the percolation threshold in a composite nanoparticle system consisting of metallic $RuO_2$ grains embedded into dielectric $CaCu_3Ti_4O_{12}$ (CCTO) matrix and annealed at $1100^{O}C$. To understand the nature of the dielectric response, we compare CCTO fabricated by two different techniques, solid state process ($CCTO_{SS}$) and sol-gel process ($CCTO_{SG}$) with the intrinsic dielectric constant in both cases found to be on the order of $10^3$-$10^4$ at 10 kHz. For $RuO_2/CCTO_{SS}$ and $RuO_2/CCTO_{SG}$ composites and composites, an increase of the dielectric constant by factors of 7 and 5 respectively is observed in the vicinity of the percolation threshold of about 0.1, with moderate losses at room temperature. Scanning electron microscopy and Energy Dispersive X-ray spectroscopy indicate that the difference in the size of the effect may arise from the microstructure of the copper oxide enriched grain boundaries in the host CCTO.

Attaining large volumetric capacities (capacities per unit volume), is of crucial importance for a number of applications, such as energy storage and renewable energy.[1] The development of double layer electrolytic capacitors, or supercapacitors, as well a number of other traditional approaches, have been successfully implemented to raise the volumetric capacity to approximately $10^2/cm^3$.[1, 2] However, in order to further increase volumetric capacity, it may be necessary to consider alternative approaches. Recently a different approach to achieve high volumetric capacity based on a rapid increase of the dielectric constant near the percolation threshold of a composite system of dielectric and metallic nanoparticles, was proposed by Efros.[3] This approach, which we will refer to as *percolative capacity*, is based on the disappearance of a continuous conduction network (infinite cluster) near the threshold, which results in the divergence of the dielectric constant $\varepsilon$, as predicted earlier by Dubrov et al.[4], Efros and Shklovskii[5], and Bergman and Imry.[6] Experimentally, this idea was successfully demonstrated in a few systems.[7, 8] However, while a large enhancement has been observed, the overall values of the dielectric constants achieved near the threshold, up to $8x10^4$, were comparable with the values attainable in a single phase ceramics, such as CCTO.

In this Letter, we show that percolative capacity can be implemented in a composite nanopaticle system consisting of insulating ceramics with very high dielectric constant, such as CCTO, and metallic oxides, such as $RuO_{2.}$ Using this system, we have achieved both high initial dielectric constant of CCTO and a significant enhancement of the dielectric capacities near the percolation threshold in the composite system using two alterative synthesis techniques for CCTO, solid state and sol-gel processes. This opens up opportunities for developing new composite materials suitable for fabricating capacitors with high dielectric permittivity.

$CaCu_3Ti_4O_{12}$ (CCTO) is a material with a very high dielectric constant, ranging from $10^3$-$10^5$ at room temperature with a quadruple perovskite structure.[9-11]. There is a significant effort being made to tune the dielectric properties of CCTO by varying the fabrication procedure, including changing the sintering temperature, time of sintering, and ambient atmosphere.[12-15]. The origin of the large, relatively temperature independent, dielectric constant in CCTO is alternatively attributed to either the intrinsic crystal structure, with the off-center displacement of Ti atoms producing local dipole moments, or to extrinsic effects such as internal barrier layer capacitance and contact – electrode depletion effect. [9, 10, 15, 16] It is also believed that the internal boundary layer capacitance effect in CCTO, in which the core grains are semi-conductive and the grain boundary is electrically insulating also attribute to high dielectric constant.[17,18] Impedance spectroscopy confirms the existence of semiconducting grains, which dominate the response at higher frequencies, together with leaky grain boundaries that are more significant at low frequencies.[19-22].

Percolative composite systems are being increasingly used in high charge storage capacitors, embedded capacitor technology, gas sensors, and in spintronics.[5, 23] Earlier an attempt was made to improve the dielectric permittivity in ceramic/polymer hybrid film that contains CCTO as functional filler in polymer host.[24, 25] Dielectric constant of 50 was obtained at low frequencies. Using CCTO as functional filler researchers have also investigated dielectric effects in two component systems, such as CCTO/Epoxy and three-component systems, such as Al/CCTO/Epoxy and Ni/CCTO/PVDF.[26-28] At room temperature, dielectric constants of 50 and 700 were reported for two- and three- component system respectively. Recently CCTO/Ag composites were prepared at 1050$^o$C to study the Maxwell-Wagner relaxation effect as well as the enhancement in permittivity in the composite system at low frequency.[29, 30] However, it was found that there is significant loss of Ag in the composite system due to the evaporation of Ag during the sintering process. The distribution of Ag was found to be inhomogeneous with high concentrations at the grain boundaries. The resulting composites showed large dielectric loss with no significant improvement of the dielectric permittivity.

The implementation of the percolation effect in CCTO is significantly complicated by the need to be annealed at least at 1100$^oC$, thus markedly restricting the number of metallic components one can use. To circumvent this problem, we have loaded dielectric CCTO matrix with metallic Ruthenium Oxide ($RuO_2$) particles. Since the sublimation temperature of $RuO_2$ is 1200K, it is considered to be stable at high temperature, and thus proves to be a good candidate for percolative composite systems. In order to better understand the nature of dielectric constant enhancement, we have prepared the host CCTO material by two different procedures, solid state process ($CCTO_{SS}$) and sol-gel (SG) process. Since the dielectric CCTO prepared by these techniques are expected to have different microstructure, comparing the dielectric properties of these composites allows us to understand the role of intrinsic and extrinsic effects.

Following the solid state process, powder samples of CCTO were prepared by mixing $CaCO_3$ (99.99%), CuO (99.99%) and $TiO_2$ (99.99%) in a planetary ball mill with acetone using zirconia balls. The mixture was then pre-calcined at 1000$^oC$ for 12 hours then mixed with a different volume fraction (f) of Ruthenium oxide. $RuO_2$ of 99.99% metal basic was obtained from Sigma Aldrich. Pallets of $RuO_2/CCTO_{SS}$ composites were prepared by using a cold pressed die with an uniaxial pressure of 1 GP, which were then calcined at 1100$^oC$ prior to dielectric measurement. To synthesize CCTO composites by Sol Gel technique we used $Ti(OBu)_4$, $Cu(CH_3COO)_2.H_2O$ and $Ca(CH_3COO)_2.H_2O$ reagents as precursors (Sigma Aldrich). These starting materials were mixed with hot glacial acetic acid and ethyl alcohol and were stirred with magnetic stirrer for 8 hrs. A xerogel was obtained at 85$^oC$. This xerogel was then pre-calcined at 1000$^oC$. The dry powder was then mixed with different volume fraction of $RuO_2$. Powder pallets were then finally calcined at 1100$^oC$ to form /$RuO_2$/$CCTO_{SG}$ composites. Silver paste as electrode was used and gold wires were connected on both sides of pallets for dielectric measurements in parallel plate geometry. Electrical properties of samples were measured by using 4284A precision LCR meter in the frequency range of 20Hz to 1MHz. Micrographs of composites were observed using a scanning tunneling microscope (JEOL, 6610 LV).

Fig.1a and Fig.1b shows SEM images of the pure CCTO samples. The grain size of both $CCTO_{SS}$ and $CCTO_{SG}$ has non-uniform distribution of 5-10μm. SEM image also reveals that the grain boundaries in $CCTO_{SS}$ are noticeably thicker than those in $CCTO_{SG}$. Energy Dispersive X-ray (EDX) analysis indicates that the compound $CaCu_3Ti_4O_{12}$ is stoichiometric. It also suggests that the grain boundaries have a higher concentration of copper oxide, while the micron-size grains are more titanium rich. Fig. 1c and Fig. 1d shows SEM images of the composite system, $RuO_2$/$CCTO_{SS}$ and $RuO_2$/$CCTO_{SG}$. EDX analysis was used to parameterize the distribution and connectivity of metallic $RuO_2$ clusters. We find the presence of strong peak of $RuO_2$ on $CCTO_{SS}$ grains, which implies that some metallic $RuO_2$ has diffused into the $CCTO_{SS}$ grain particles, while for the $CCTO_{SG}$ composite, the peak appears to be quite weak, indicating little diffusion into the $CCTO_{SG}$ grains. The significant diffusion of $RuO_2$ in $CCTO_{SS}$ is believed to be due to presence of high fraction of molten copper oxide grain boundary which helped $RuO_2$ to mobilize more easily whereas in $CCTO_{SG}$, thin layer of molten CuO, leads to weak diffusion of $RuO_2$. This diffusion can affect the electrical conduction across the interface in the matrix, which in turn, will have different dielectric response among two different CCTO hosts. For as prepared $CCTO_{SS}$/$CCTO_{SG}$ samples, the temperature dependence of the dielectric constant and the dissipation factor at a frequency of 30 kHz are shown in Fig. 2a and Fig. 2b respectively. Lowering the temperature increases the characteristics relaxation time of dipole moments which results in slower polarization process.[7] The plot also reveals that the dielectric constant shows two gradual drops coincident with peaks in the dielectric loss. This confirms that there are two separate relaxation mechanisms in these samples, consistent with other studies on CCTO.[11,24] Fig.3 shows the dependence of dielectric constant at 30 kHz as a function of volume fraction of $RuO_2$ in $CCTO_{SS}$ ($CCTO_{SG}$)/$RuO_2$ composite systems. Both series of composites show a dramatic increase in the dielectric constant at high frequency near a certain composition, consistent with percolation effects. At low frequency, Maxwell-Wagner effect dominates, as this arises from the space charge polarization and shows a strong dependence of the dielectric response in percolation system. Due to this reason, the dielectric response at intermediate frequencies is studied. At 30 kHz, the dielectric constant $\varepsilon_{eff}$ for as prepared $CCTO_{SS}$ and $CCTO_{SG}$ is found to be approximately $4x10^3$ and $3.3x10^3$ at room temperature respectively. In the $RuO_2$/$CCTO_{SS}$ composite, the dielectric constant $\varepsilon_{eff}$ increases by approximately by a factor of 7 at a 11% volume fraction of $RuO_2$ ($f_c = 0.11$), while for $RuO_2$/$CCTO_{SG}$, $\varepsilon_{eff}$ increases by a factor of 5 at 8% volume fraction ($f_c$=0.08). The plot reveals the percolation-type behavior as the dielectric constant tends to diverge at specific metallic volume fractions. The main contribution for the enhancement of the dielectric constant comes from the formation of microcapacitor networks (the metallic clusters separated by layer of polarized dielectric) in the composite systems.[31-33,5]. This result suggests that dielectric constant of CCTO can be improved even when metallic particles are incorporated into the host CCTO. This is completely different from other experiments, in which CCTO was used as metallic fillers in host polymers.[24,26] The increase in the dielectric constant of the composite is found to be frequency independent. The effective dielectric constant $\varepsilon_{eff}$ follows a power law scaling as a function of $RuO_2$ composition and is given as $\varepsilon_{eff} \propto (f_c - f)^{-q}$, for $f < f_c$, and $(f - f_c)^{-t}$ for $f > f_c$, where $f$ is the volume fraction of ruthenium oxide, $f_c$ is the percolation threshold, and $q$ and $t$ are the critical exponents. The inset in Fig. 3a and Fig. 3b shows that for $RuO_2/CCTO_{SS}$ and $RuO_2/CCTO_{SG}$ composites, the values found for $f_c$ are 0.11 and 0.08 respectively, and the values of $q$ are 0.3 and 0.2 respectively. The low percolation thresholds in our composites is due to the finite conductivity of the grains and the grains boundaries in the dielectric CCTO resulted from interfacial $RuO_2$ diffusion, which leads to deviation from the universal threshold value of 0.16.[34] Due to the same reasons, the critical exponents of the dielectric constant $q$ = 0.3 and 0.2 for our composites also differ substantially from the universal value of 0.8-1.[35] We have also determined the critical exponent $t$ on the metal side of the metal-insulator transition $\varepsilon_{eff} \propto (f - f_c)^{-t}$, which is found to be about 0.15 and 0.06 for $RuO_2/CCTO_{SS}$ and $RuO_2/CCTO_{SG}$ respectively. Similar to the change of dielectric constant with $f$, the dielectric loss ($D$) of the composite also increases with $f$, as shown in fig. 4a. For $RuO_2/CCTO_{SS}$ composites, it is found that at 30 kHz, the loss $D < 2$ for $f < f_c$ (0.11) and for $RuO_2/CCTO_{SG}$ composites, $D < 1.5$ for $f < f_c$ (0.08). Above the percolation thresholds, the dielectric loss increases significantly due to increase of leakage current. Similarly as shown in fig. 4 (b), the increase of conductivity ($\sigma$) at the vicinity of $f$ = 0.11 and 0.08 is clearly seen. The best fit of the conductivity data of $RuO_2/CCTO_{SS}$ and $RuO_2/CCTO_{SG}$ composites to the equation $\sigma \propto (f_c - f)^{-q}$, for $f < f_c$. yields $q$ = 0.4 and $q$ = 0.27 respectively. We note that the values of $q$ found from the electrical conductance transition are proportionally higher, compared to the ones found from the dielectric power law shown in Fig. 3. This is likely to reflect the complex nature of the percolative transition, which will be addressed in more detail elsewhere. Accordingly, the conductivity of the dielectric composites increases with frequency as shown in Fig. 4(c-d). It also shows that for $f < f_c$, the composites have low conductivity but at percolation threshold, there is a jump in conductivity indicating high leakage current. This same behavior is found for the dielectric loss as $f \rightarrow f_c$.

In summary, we have fabricated CCTO by both solid state and sol-gel processes with high values of the dielectric permittivity. By introducing $RuO_2$ in the dielectric matrix we were able to further increase the dielectric constant in the vicinity of the percolation threshold by about 5-10 times. In both $RuO_2/CCTO_{SS}$ and $RuO_2/CCTO_{SG}$ composites, the dielectric response is found to be frequency independent. The values of the critical exponent $q$ for our composite system were found to be lower than in other systems, which can be attributed to different types of interaction between semiconductor (CCTO) and metallic particles ($RuO_2$). While we were generally successful in using $RuO_2$ as a metallic component, some diffusion of $RuO_2$ into the CCTO matrix has occurred, particularly in the case of $CCTO_{SS}$, which was likely to limit the amount of increase of the dielectric constant we observed, as it would broaden the percolative transition. One can further optimize the fabrication procedure to control the size and composition of grain boundaries, the degree of inter-diffusion, as well as the size of CCTO nanoparticles, which should be much smaller (about 50-100 nm) to achieve the best results. From this perspective our current work is a proof of concept, as the increase of the dielectric constant near the percolation threshold is fairly modest compared to other systems. By the same token, however, that indicates that there is a lot of room for improvement in this approach, so that it can potentially lead to dielectric constants in the $10^6$ range, which would important implications for various applications. This work was supported by the National Science Foundation under DMR-1006381.

Figure Captions.

Fig.1. SEM image shown for (a) pure $CCTO_{SS}$ made by solid state process, (b) pure $CCTO_{SG}$ made by sol gel process, (c) $RuO_2$/ $CCTO_{SS}$ composite and (d) $RuO_2/CCTO_{SG}$ composite. $RuO_2$ particles are represented by arrows.

Fig.2. The variation of dielectric constant with temperature is shown for (a) $CCTO_{SS}$ and (b) $CCTO_{SG}$ at 30KHz. The insets show the temperature dependency of dielectric loss corresponding to the same frequency.

Fig.3. Enhancement of dielectric constant with the volume fraction of $RuO_2$ ($f$) is shown for (a) $RuO_2/CCTO_{SS}$ composites and (b) $RuO_2/CCTO_{SG}$ composites. Insets: log-log fit for determining the critical exponents below and above $f_c$.

Fig.4. (a) Variation of dielectric loss ($D$) as a function of $f$ in $RuO_2/CCTO_{SS}$ and $RuO_2/CCTO_{SG}$ composites, (b) conductivity as a function of $f$ at 30KHz for both the composites; the insets shows the power law behavior at percolation threshold, (c) conductivity as a function of frequency for series of $RuO_2/CCTO_{SS}$ composites with increasing $f$, (d) conductivity as a function of frequency for $RuO_2/CCTO_{SG}$ composites.

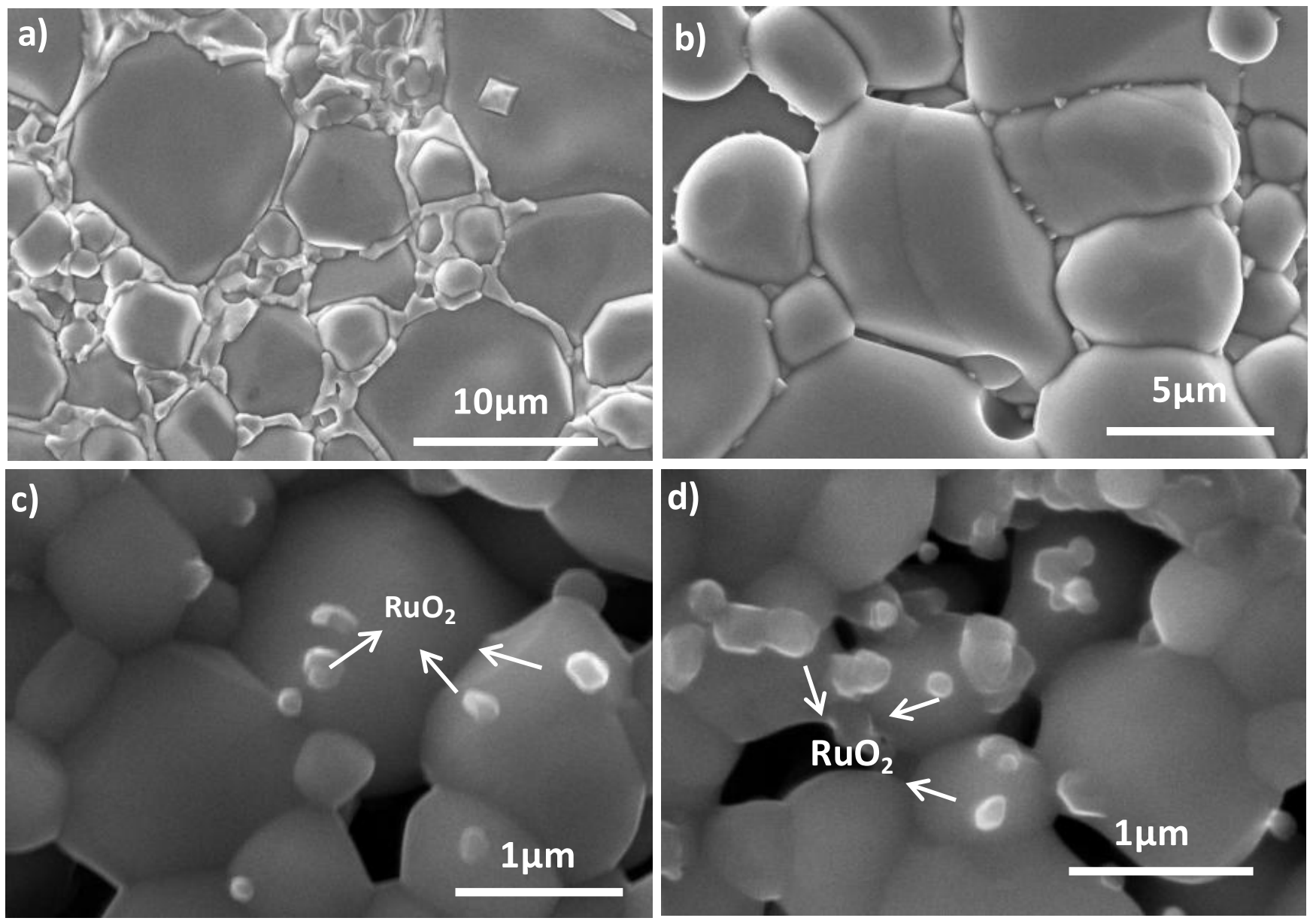

Fig.1

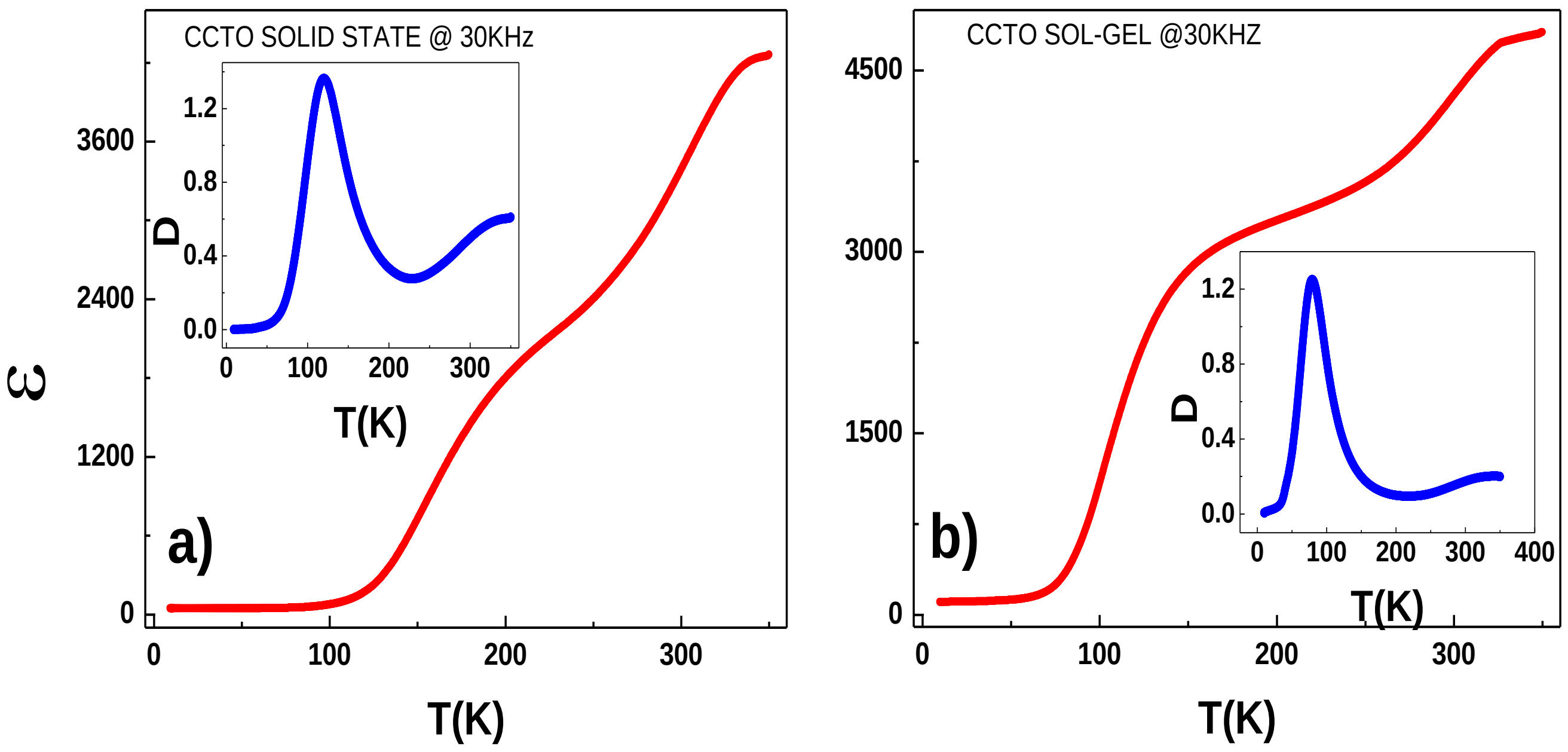

Fig.2

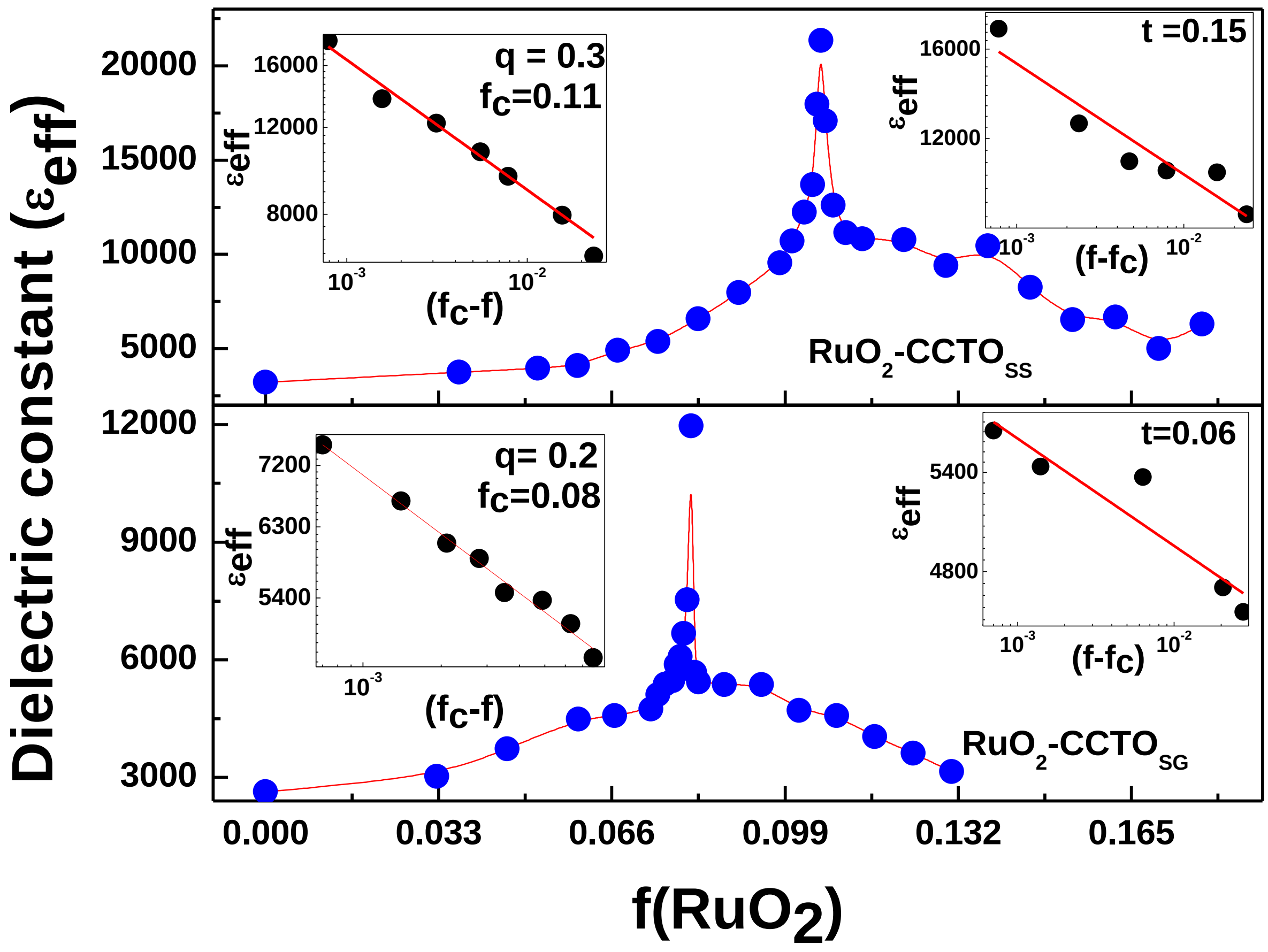

Fig.3

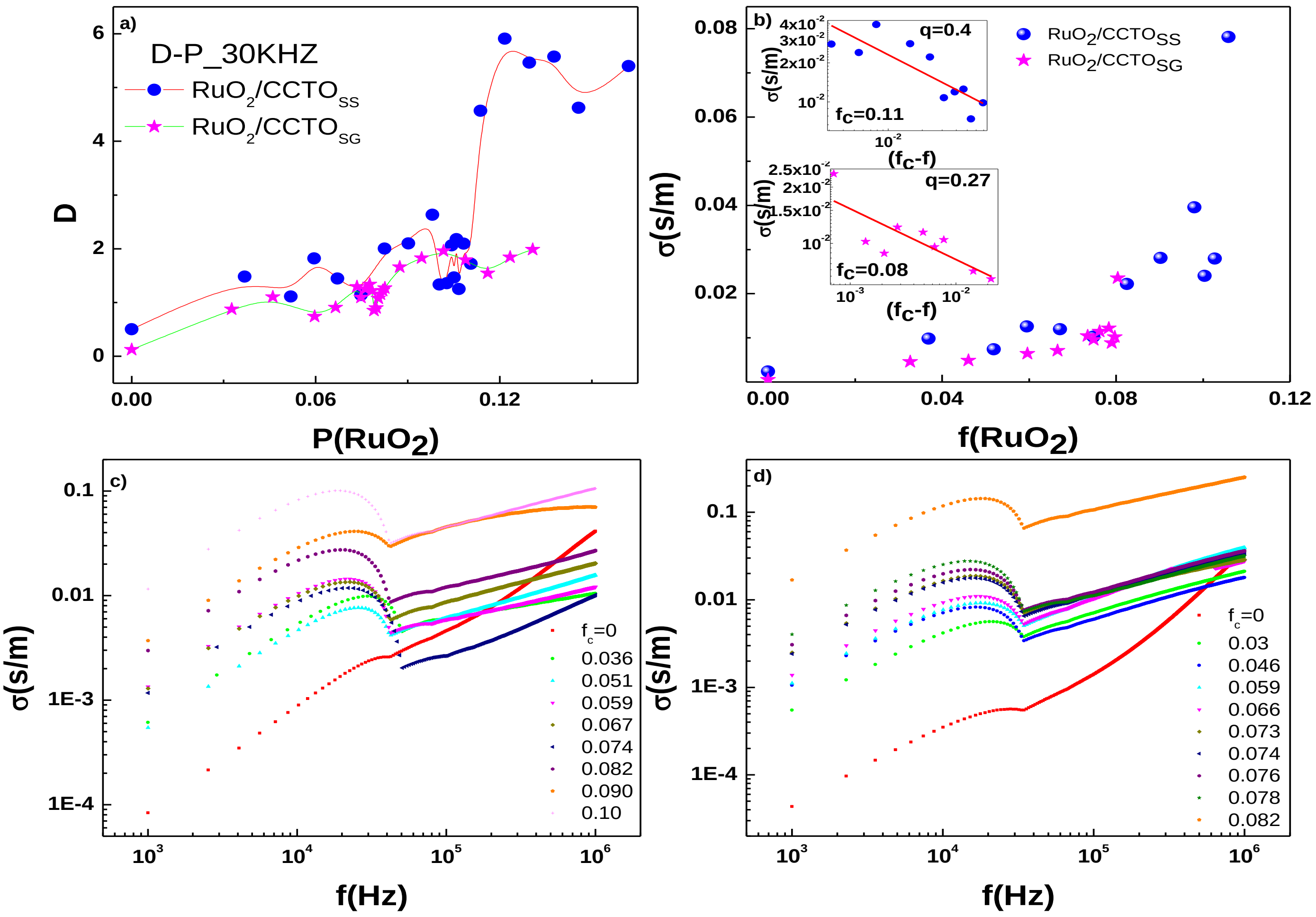

Fig.4